\documentclass[pra,showpacs,onecolumn,byrevtex,superscriptaddress]{revtex4}
\usepackage{graphicx}
\usepackage{subfigure}
\usepackage{dcolumn}
\usepackage{amsmath}
\usepackage{epsfig}
\usepackage{color}
\usepackage{amsmath}
\usepackage{multirow}
\usepackage{caption}

\begin{document}
\title{Simple security proofs for continuous variable   quantum key distribution with intensity fluctuating sources}

\author{Chenyang Li}
\email{chenyangli@ece.utoronto.ca}
\affiliation{Center for Quantum Information and Quantum Control, Department of Electrical \& Computer Engineering,University of Toronto,
Toronto, M5S 3G4, Canada}
\affiliation{Department of
Physics, The University of Hong Kong, Hong Kong, China}

\author{Li Qian}
\affiliation{Center for Quantum Information and Quantum Control, Department of Electrical \& Computer Engineering,University of Toronto,
Toronto, M5S 3G4, Canada}

\author{Hoi-Kwong Lo}
\affiliation{Center for Quantum Information and Quantum Control, Department of Electrical \& Computer Engineering,University of Toronto,
Toronto, M5S 3G4, Canada}
\affiliation{Department of Physics, The University of Hong Kong, Hong Kong, China}
\affiliation{Department of Physics, University of Toronto, Toronto, M5S 3G4, Canada}

\begin{abstract}
  Despite tremendous theoretical and experimental progress in continuous variable (CV) quantum key distribution (QKD), the security has not been rigorously established for most current continuous variable quantum key distribution systems that have imperfections.  Among these imperfections, intensity fluctuation is one of the principal problem affecting  security. In this paper, we provide  simple security proofs for continuous variable quantum key distribution systems with intensity fluctuating sources.  Specifically, depending on device assumptions in the source, the imperfect systems are divided into two general cases for security proofs. In the most conservative case, we prove the security based on the tagging idea, which is a main technique for the security proof of discrete variable quantum key distribution. Our proofs are simple to implement without any hardware adjustment for  current continuous variable quantum key distribution systems. Also, we show that our proofs are able to provide secure secret keys in the finite-size scenario.
\end{abstract}

\maketitle

\section{Introduction}
\normalsize

Quantum key distribution (QKD) allows two distant parties to share a common string of secret data \cite{Lo2014,Weedbrook2012,Diamanti2015}. Based on the laws of quantum mechanics, QKD offers information-theoretical security.  QKD has aroused much interest in both theoretical protocol and experimental demonstration, because it is considered the first application of quantum information science to reach commercial maturity. For example, the implementation of discrete variable (DV) QKD protocols including
satellite-to-ground QKD \cite{Liao2017} and chip-based QKD \cite{Ma2016,Sibson2017,Li2017} have demonstrated the potential for commercial applications in the filed of quantum information. Besides, twin-field QKD\cite{Lucamarini2018,Xiaoqing2019} has been proposed to outperform the well-known rate-loss limit \cite{Pirandola2017} and largely extend  transmission limits. Compared to DV protocols, continuous variable (CV) protocols have the potential for high-key rate and low-cost implementations using current standard telecom components such as homodyne detectors \cite{Diamanti2015}. Recently, CV QKD experiment has demonstrated the secret key transmission over a long distance  from 100 km \cite{zeng2016}  to more than 200 km\cite{Hong2020}.

Despite the enormous progress in the field of QKD,  the most important question in quantum communication is always how secure QKD really is. For example, are QKD systems secure when implemented with practical devices? Fortunately, measurement-device-independent QKD \cite{Lo2012} can remove all imperfections and security loopholes in
the measurement devices, and therefore we only need to consider the imperfections in the source devices.
Imperfect sources, such as the correlated intensity fluctuations in optical pulses \cite{Yoshino2018} and setting-choice-independently correlated light sources \cite{Mizutani2019}, have been recently analyzed in DV QKD systems.  However, the security research concerning CV QKD with imperfect source has fallen behind that of its discrete-variable cousin.
For instance, almost all existing CV QKD proofs require a perfect state preparation \cite{Jouguet2013}, i.e., Gaussian modulation, which cannot be guaranteed in a practical CV QKD system with imperfections and limitations\cite{Jouguet2012,Wenyuan2017}.
The security of continuous-variable quantum key distribution with noisy coherent states has been analyzed in \cite{Filip2008,Usenko2010,Shen2011} by introducing an independent and additive Gaussian noise to a perfect Gaussian modulation.
However, in the practical continuous variable modulation, the imperfections might not work independently or additively with Gaussian modulation. For example, intensity fluctuation is one of the potential practical problems affecting the use of Gaussian modulation by its dependence on modulated quadratures.
Therefore, in this work we study intensity fluctuations in practical CV QKD systems.  Our intensity fluctuation model is an arbitrary distributed random variable with a unit mean value. Depending on whether the intensity fluctuation information is accessible or not to Alice, our security analysis of a QKD system can be generally divided into two cases:(1)  Alice  can ,and (2) Alice cannot monitor intensity fluctuation values for every pulse.

In this work, we prove the security for the two cases based on different techniques.  Particularly, in case (1) , because Alice's information can help modify her data, the security proof is based on the integrating over the distribution of intensity fluctuations.  Also, a refined data analysis is developed to improve the QKD performance over long distance.  In case (2), Alice can not exactly monitor  every signal pulse. Depending on whether Eve has the intensity fluctuation  information, we divide  case (2) into two subcases: (2A)Eve can, and (2B)Eve cannot monitor intensity fluctuation values for every pulse. In subcase (2A), we prove the security based on Gaussian extremality\cite{Michael2006,Raul2006}. In the most conservative case (2B),  we apply the concept of tagging, previously developed for DV QKD in \cite{Gottesman2002} , to the security proof of CV QKD. Specifically, we divide up signals into two distinct sets, untagged and tagged. Untagged signals are those whose intensities fall inside a prescribed region, whereas tagged signals are those whose intensities might fall outside the prescribed region. In the actual protocol, the QKD system users do not need to know whether each signal is tagged or untagged. They only need to be able to  set a bound for untagged signals, which would lead to the security of their generated key. Moreover, given the distribution of intensity fluctuations, the users could obtain the probability of untagged signals and further optimize the secret key rate by the fraction of untagged signals. In the end, our proofs for all cases are simple to implement without any hardware adjustment for the current continuous variable quantum key distribution system. Alice and Bob are free to choose choose different security proofs to generate the secret key based on their device assumptions. In the end, we demonstrate that our proofs  are able to provide secure secret keys in the finite-size scenario over distances larger than 50 km.

\section{Results}

\textbf{Intensity Fluctuation Model}

Here, we  define our  model for experimental intensity fluctuations.  For example, suppose that a desired pulse intensity  is $I_A$, however, Alice actually prepares a pulse with the intensity of $kI_A$.  We  denote $k$ as a random variable to characterize the intensity fluctuations, with mean value $E_k$ and variance $V_k$. This intensity fluctuation can be caused
by power fluctuations of a  laser or imperfect intensity  modulators \cite{Laudenbach2018}.    In this paper, for simplicity, we assume the following conditions of the random variable $k$:

1) $k$ is an independent and identically distributed (i.i.d.)   random variable.

2) $k$ has a mean value $E_k$ and a variance $V_k$, where $E_k$ is 1.

3) $k$ is independent of the pulse intensity $I_A$.

4) the probability distribution function of $k$ can be obtained before the experiment by testing the source device.

5) the probability distribution function of $k$ will not change during the QKD transmission.

Here, these conditions are assumed to simplify our model for experimental intensity fluctuations. Conditions 1)-3) are the intrinsic constraints and assumptions for the intensity fluctuations.
Conditions 4)-5) are the assumptions for system characterization, which is required before QKD transmission.

\textbf{CV QKD system description}

 Fig 1  shows that, with the intensity fluctuation information, QKD systems can be generally divided into two cases for security proofs. To fairly compare the results, an ideal CV QKD system is added as the baseline case (0) for benchmarking. Here,
  following  \cite{Gottesman2002} we introduce a hypothetical party Fred, who  controls the intensity fluctuations $k$ for every optical pulse, e.g., the intensity fluctuation can be controlled by temperature drift. Through secure communication, Fred would choose to reveal the value of $k$  to Alice.   In total, there are two cases:

(1) Fred discloses the actual value of $k$ to Alice;

(2) Fred does not disclose the actual value of $k$  to Alice.

In both cases, because the actual pulse intensity is $kI_A$, the actual encoded Gaussian random variable now becomes $\sqrt{k}X_A$ and Alice sends out a  mode $\hat{A}_1=\hat{0}+\sqrt{k}X_A $.  In case (1),  Alice has access to the intensity fluctuation values $k$ and can further revise her data from $X_A$ to $\sqrt{k}X_A$ for every pulse.  In case (2) , Alice does not have access to the intensity fluctuation values $k$. Depending on whether Eve has the intensity fluctuation side information, we divide  case (2) into two subcases (2A) and (2B) for security proofs.

For a common QKD system, it is usually assumed that Eve often has infinite power in the channel with only limitations from the laws of physics. In other words, the source should always be assumed to be secure and  no information in the source stage can be disclosed to Eve. Here, we divide the QKD systems into different cases only based on the source information leakage assumptions. It is open for Alice and Bob to consider which case is acceptable in their QKD transmission process. For case (1), the justification is that Alice can have access to the device imperfection in real time. For case (2A), the justification is that Alice should use a certified device which come from a faithful company. For case (2B), this is most conservative case. If Alice does not have enough confidence on the device, they can always choose case (2B).  Note that, the authors in \cite{Namiki2018} have  applied the similar idea to the detection stage where they assume that the detection process is inaccessible to eavesdroppers.

\begin{figure*}[!htb]
\begin{minipage}{0.4\linewidth}
\centerline{\includegraphics[width=1\textwidth]{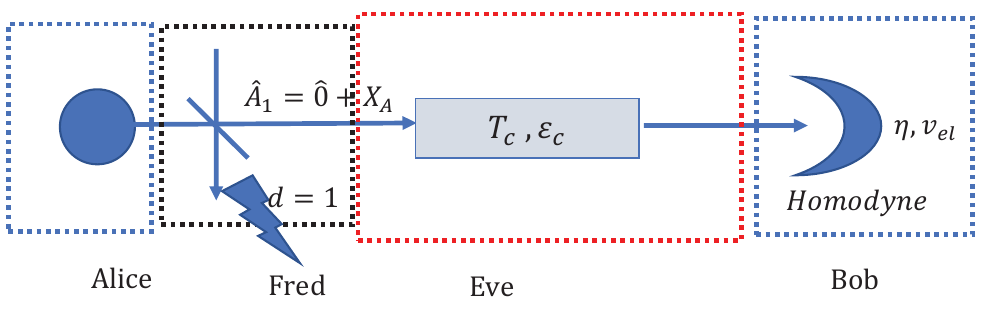}}
\centerline{(0) ideal CVQKD system ($k=1$)}
\end{minipage}
\\
\begin{minipage}{0.4\linewidth}
\centerline{\includegraphics[width=1\textwidth]{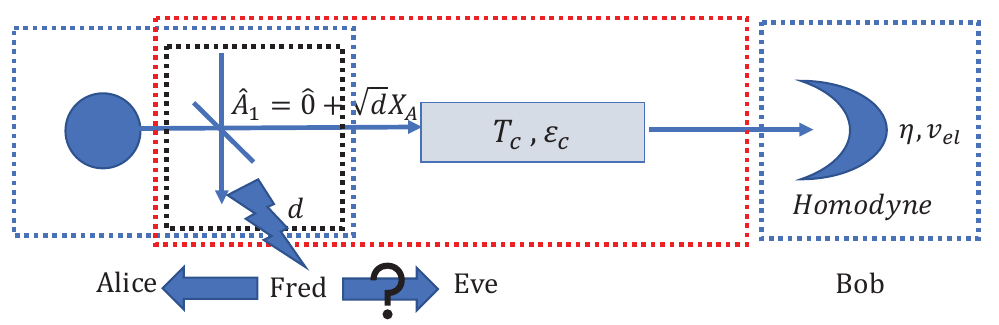}}
\centerline{(1) $k$ is disclosed to Alice}
\end{minipage}
\qquad
\begin{minipage}{0.4\linewidth}
\centerline{\includegraphics[width=1\textwidth]{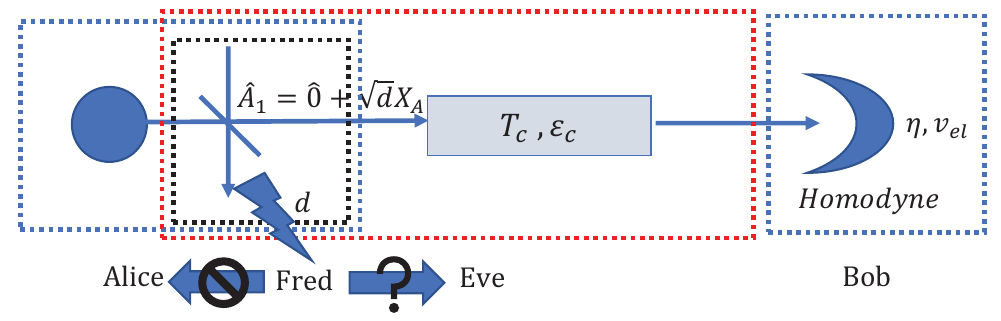}}
\centerline{(2) $k$ is not disclosed to Alice }
\end{minipage}

\caption{Here, practical CV QKD systems can be divided into two cases based on the Alice's information about  intensity fluctuations. One ideal case (0) is added for comparison. In case (0), a CV QKD system does not have  any intensity fluctuations. In case (1), Alice can monitor the intensity fluctuations.   In case (2), Alice cannot  monitor the intensity fluctuations. Depending on whether Eve has  intensity fluctuation information or not, case (2) is divided into two subcases (2A) and (2B).  Here, $T_c$ and $ \varepsilon_c$ are, respectively, the channel transmittance and excess noise between Alice and Bob.  $\eta$ and $v_{el}$ are the the detection efficiency and electronic noise of the homodyne detector. Here, the symbol "?" in case (1) means two possible subcases that Eve can or cannot have access to the intensity fluctuation information.   The No Entry sign in case (2) means that the intensity fluctuation information will not be disclosed to Alice or Eve. }
\end{figure*}

~\\

\textbf{Security proof for case (0)}

Here, we briefly review the security proof for an ideal CV QKD system.  Because the security against coherent attacks can be reduced to that against collective attacks by using de Finetti representation theorem  for infinite dimensions \cite{Renner2009}, for simplicity, we only consider asymptotic security against collective attack . Given reverse reconciliation communication,   the asymptotic secret key rate is given by the Devetak-Winter formula \cite{Devetak2005,Lodewyck2007,Jouguet2011}:
 \begin{gather}\label{eq1}
    R_0 = \beta I_{AB}-\chi_{BE}
\end{gather}
where $\beta$ is the reverse reconciliation efficiency,  $I_{AB}$ is the mutual information between Alice and Bob, and $\chi_{BE}$ is the mutual Holevo information between Bob and Eve.
Given parameter estimations of  transmittance $T$ and excess noise $\varepsilon$, the computation for $I_{AB}$ and $\chi_{BE}$ can be found in the Supplementary Section I.

~\\

\textbf{Security proof for case (1)}

In case (1), Alice has access to the intensity fluctuation values $k$ and can further revise her data from $X_A$ to $\sqrt{k}X_A$ for each pulse.  The security proof is based on two conclusions: a) the strong superadditivity of secret key rate; b) the weak law of large numbers.

Suppose Alice and Bob share $n$ modes in a joint state $\rho_{A_{1,2,...n}B_{1,2,...n}}$, and Alice has the intensity fluctuation information $k_i$ for the $i^{th}$ mode.  Conditional on the $k_i$, The secret key rate for this joint state can be shown as
 \begin{gather}\label{eq1}
    R_1 = \frac{1}{n} R(\rho_{A_{1,2,...n}B_{1,2,...n}|k_1k_2...k_n})\geq  \frac{1}{n} \sum_{i=1}^{n}R(\rho_{A_{i}B_{i}|k_{i}}) \\ \notag
         \rightarrow E[R(\rho_{A_{i}B_{i}|k_{i}})]= \int_{-\infty}^{+\infty}PDF(k)R_0(k,T)dk
\end{gather}
where $PDF(k)$ is the probability density function of $k$, $R(\rho_{A_{i}B_{i}|k_{i}})$ is the secret key rate conditional on the $k_i$.

In  the first line of Eq.(2) , we use the strong superadditivity of the secret key rate from \cite{Raul2006}.  Then in second line, we argue that by the weak law of large numbers,  the sum over all reduced modes converges to the average over its probability density function in the limit $n \rightarrow \infty$.

Given the intensity fluctuation information, we propose that a simple refined data analysis can be adopted by Alice to improve the maximum distance and defend against possible attacks based on intensity fluctuations.
Here, we describe a refined data analysis process  as below:
(1) Based on the probability density function of $k$, Alice will divide  $k$ into a number of sets with equal probability.
(2) Alice and Bob will perform the parameter estimation individually for each set, obtaining the channel transmittance and  excess noise and verifying whether the channel transmittance matches with that from another set.  This process is used to defend any possible attack for Eve based on intensity fluctuation information.
(3) For certain sets, if $R_0(k,T)<0$, Alice and Bob will simply drop all the data from such sets.

After a refined data analysis, the secret key rate can be expressed as
\begin{gather}\label{eq1}
    R_{1R} = \int_{-\infty}^{+\infty}PDF(k)\max\{R_0(k,T),0\}dk
\end{gather}
\begin{table}[!htb]
\centering
 \fontsize{12}{12}\selectfont
\caption{Evaluation parameters for fiber-based QKD \cite{Lodewyck2007,Jouguet2014}}
\begin{tabular}{|c|c|c|c|c|}
\hline
 $\eta $      & $ \varepsilon_c $  & $v_{el}$ & $V_A $ & $\beta $  \\
\hline
0.60      &     0.02    & 0.02 &  18 & 95.6\\
\hline

\end{tabular}
\label{table1}
\end{table}

Fig 2 shows the simulation result for the  secret key rate  $R_{0}$,   $R_{1}$ and $R_{1R}$.   We use the parameters listed in Table I, where  $\eta$ and $v_{el}$ are, respectively,  the detection efficiency and electronic noise of the homodyne detector, $\varepsilon_c$ is the excess noise in the channel, $V_A$ is the modulation variance and $\beta$ is the reverse reconciliation efficiency.
In Fig 2(a), we  choose the  probability density function of $k$ to be an uniform distribution from $0.9$ to $1.1$. In Fig 2 (b), we  choose the  probability density function of $k$ to be an uniform distribution from $0.8$ to $1.2$.  Through simulation,  we find that the secret key rate $R_{1}$ is approximately same as  $R_0$. By refined data analysis, the maximum  transmission distance  can be improved from  94km to 130km in Fig 2(a), and from 94 km to  199km in Fig 2(b).  This maximum transmission distance improvement is expected, since the refined data  analysis can be regarding as a  pre-selection of    optimal Gaussian states for long distance.

\begin{figure*}[!htb]

\subfigure[Uniform distribution from 0.9 to 1.1 for $R_1$ ]{
\includegraphics [width=75mm,height=60mm]{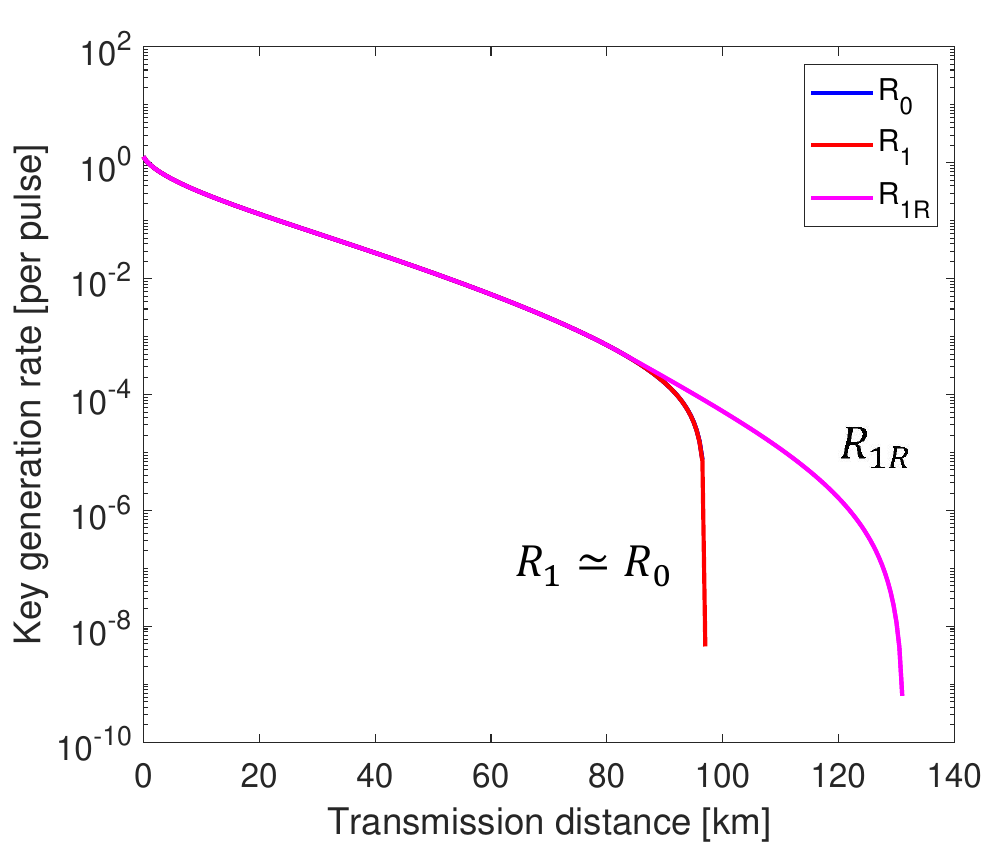}}
\subfigure[Uniform distribution from 0.8 to 1.2 for $R_1$]{
\includegraphics [width=75mm,height=60mm]{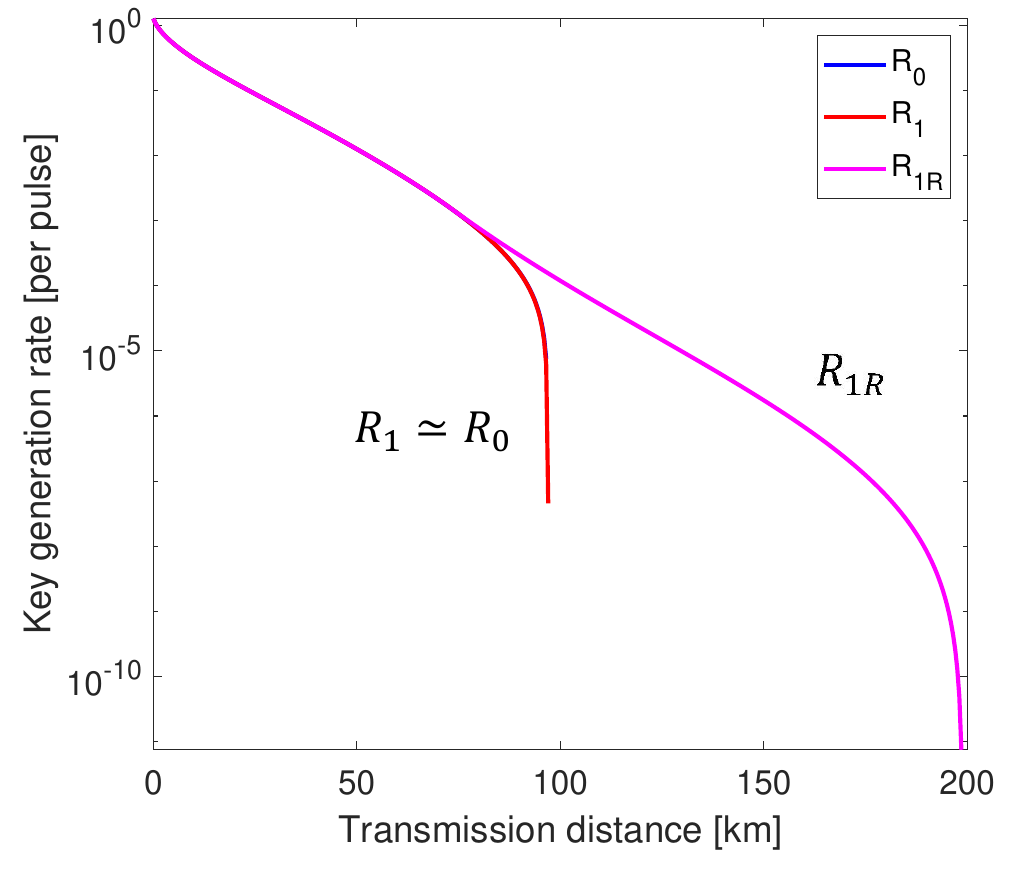}}

\caption{Here, we compare the secret key rate, $R_{0}, R_1$ and  $R_{1R}$. In Fig 2(a), the intensity fluctuation model is a uniform distribution from 0.9 to 1.1. The secret key rate $R_{1}$ is approximately same as the key rate $R_0$ for ideal CV QKD system. In Fig 2(b), the intensity fluctuation model is a uniform distribution from 0.8 to 1.2.  It is clearly demonstrated that  both maximum transmission distances  can be improved by refined data analysis.}
\end{figure*}

~\\

\textbf{Security proof for case (2A)}

\begin{figure*}[!htb]

\centering
\includegraphics [width=120mm,height=40mm]{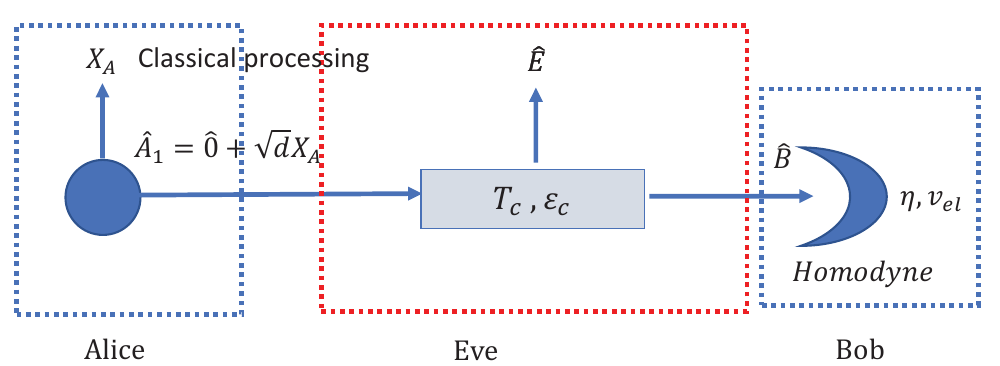}

\caption{Here, we consider case (2A) that Eve also has no intensity fluctuation information. Therefore, Eve can only manipulate the signal states in the channel. Due to intensity fluctuation, Alice will have a recorded data mismatched with what she really encodes.}
\end{figure*}

Here, we consider case (2A): Alice and  Eve both  have no intensity fluctuation information. As shown in Fig 3, for each pulse, Alice has no intensity fluctuation information and can only record the data $X_A$. However, what Alice really encodes is the mode $\hat{A}_1=\hat{0}+\sqrt{k}X_A$.    By considering reverse reconciliation  with the Bob's recorded data $X_B$, The secret key rate can be expressed as
 \begin{gather}\label{eq1}
    R_{2A} = \beta I(X_A,X_B)-\chi(X_B,E)_{|\sqrt{k}X_A}
\end{gather}
where $I(X_A,X_B)$ is the mutual information between Alice's and Bob's classical recorded data $X_A$ and $X_B$, and $\chi(X_B,E)_{|\sqrt{k}X_A}$ is the Holevo mutual information between  Bob and Eve given the actual input mode $\hat{A}_1$ before the channel. Here, $I(X_A,X_B)$ can be directly obtained from the data sets, while an upper bound for $\chi(X_B,E)_{|\sqrt{k}X_A}$ is needed.
Next, we use the Gaussian extremality \cite{Michael2006,Raul2006}  that the Holevo information $\chi(X_B,E)_{|\sqrt{k}X_A}$ between Eve's and Bob's classical variables, is maximized when then the state $\rho_{AB}$ shared by Alice and Bob is Gaussian. In other words, we can obtain the upper bound of $\chi(X_B,E)_{|\sqrt{k}X_A}$   by substituting Alice's and Bob's actual mode $\hat{A}_1, \hat{B}$ with  Gaussian modes which have the same first and second  quadrature moments.
By calculating the mean value and variance of $\sqrt{k}X_A$, we can obtain that $<\sqrt{k}X_A>=<X_A>=0, <kX^2_A>=<X^2_A>=V_A$.
Furthermore, we obtain the upper bound that
\begin{gather}\label{eq1}
\chi(X_B,E)_{|\sqrt{k}X_A} \leq \chi(X_B^G,E)_{|X_A^G}
\end{gather}
where $X_A^G$ and $X_B^G$ are, respectively, the Gaussian random variable with the same first and second moments as $X_A$ and $X_B$.

Next, we will estimate  the equivalent transmittance $T_s$ and excess noise $ \varepsilon_s$  in the source caused by the data mismatch. According to the Supplementary Section II, suppose Alice  records $X_A$ and the actual encoded data is $\sqrt{k}X_A$,  the equivalent  $T_s$ and $ \varepsilon_s$  can be expressed as
\begin{gather}
T_s=<\sqrt{k}>^2 \simeq (1-\frac{1}{8}V_k)^2, \\  \notag
\varepsilon_s=\frac{V_A}{T_s}-V_A \simeq \frac{1}{4}V_AV_k,
\end{gather}

In addition to the channel transmittance $T_c$ and excess noise $\varepsilon_c$, Alice and Bob would estimate an overall transmittance $T$ and  excess noise $\varepsilon$ such that
\begin{gather}\label{1}
  T=T_sT_c,           \\  \notag
  \varepsilon=\varepsilon_c/T_s+\varepsilon_s
\end{gather}

\begin{figure}[!htb]
\centering
\subfigure[Uniform distribution  ]{
\includegraphics [width=75mm,height=55mm]{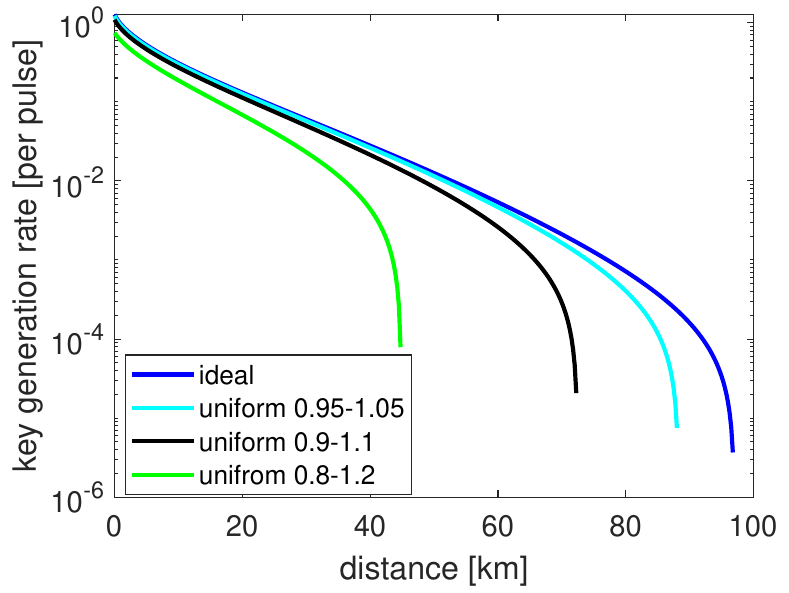}}
\subfigure[Gaussian distribution ]{
\includegraphics [width=75mm,height=55mm]{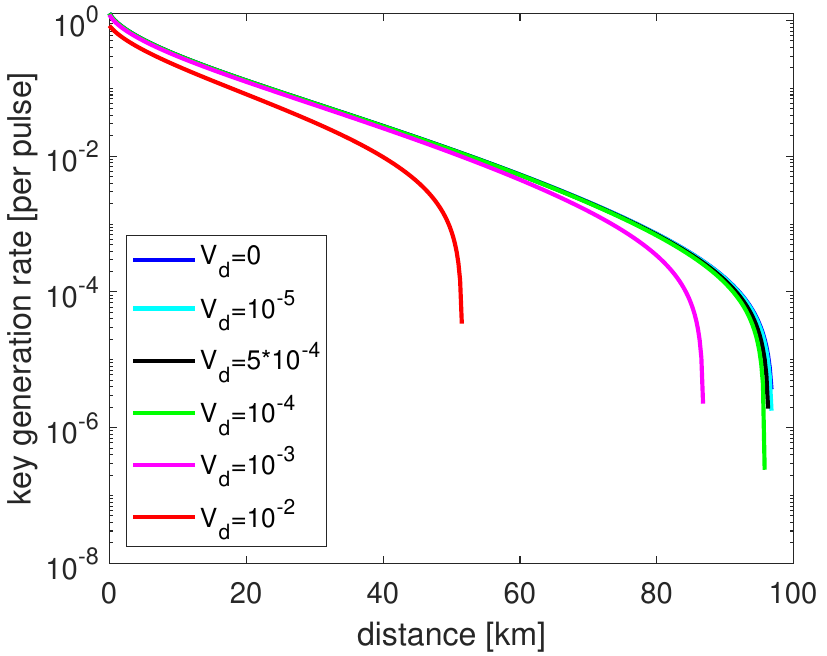}}

\caption{Here, we compute the secret key rates $R_{2A}$ with two  intensity fluctuation models. (a)The secret key rates versus transmission distance for different intensity fluctuation models of uniform distribution.  (b)The secret key rates versus transmission distance for different intensity fluctuation models of Gaussian distribution. }
\end{figure}\label{idac}

Fig 4 shows the secret key rate for case (2A). We still use the channel and detector parameters listed in Table I. In Fig  4(a),  we compute the secret key rates  for the  uniform distributed intensity.  Even if  the pulse intensity  fluctuate  5\%, the maximum transmission distance will still drop about 10 km.
In Fig  4(b),  the secret key rates are obtained for the  Gaussian distributed intensity.  The variances of the Gaussian distribution range from $0$ to $10^{-2}$. When the variance increases to $10^{-2}$, the maximum transmission distance will decrease by  about 40 km. In other words, when the standard deviation  of Gaussian distribution is $10\%$, the maximum transmission distance will drop significantly.

~\\

\textbf{Security proof for case (2B)}

In this section, we consider case (2B): Eve has  intensity fluctuation information while Alice has no information. Before we jump into security proof, we first define the untagged Gaussian state.  Here, we apply the concept of "tagging"\cite{Gottesman2002} to  case (2B) of CV QKD.  Suppose Alice sends out $n$ Gaussian modulated coherent pulses to Bob and the $i_{th}$ pulse has a intensity fluctuation value $k_i$.    However, Alice has no information about the intensity fluctuation value for each pulse, and  Alice can only  record data set as $k_i=1$. Now we define  the Gaussian modulated coherent states with intensity fluctuation value $k_i<1$ as  untagged Gaussian states.
It is easy to verify that when Alice sends out a stronger pulse than what she is supposed to send, Alice and Bob will definitely overestimate the secret key rate by underestimating the channel loss and excess noise.  Therefore, the untagged Gaussian states are defined to be the states from which Alice and Bob will not overestimate the secret key rate. In other words, the untagged Gaussian states are always conservative secure. Next, we can introduce an cutoff $k_{max}$ based on the intensity fluctuation probability density function. As depicted in Fig 5,
if Alice chooses a cutoff $k_{max}$, the Gaussian states associated with  lower intensities than $k_{max}I_A$  would always be untagged. Then the probability to get untagged Gaussian states can be expressed as
\begin{gather}\label{1}
    p_s=\int_{-\infty}^{k_{max}}\textrm{PDF}(k)dk
\end{gather}

\begin{figure}[!htb]
\centering
\includegraphics [width=60mm,height=50mm]{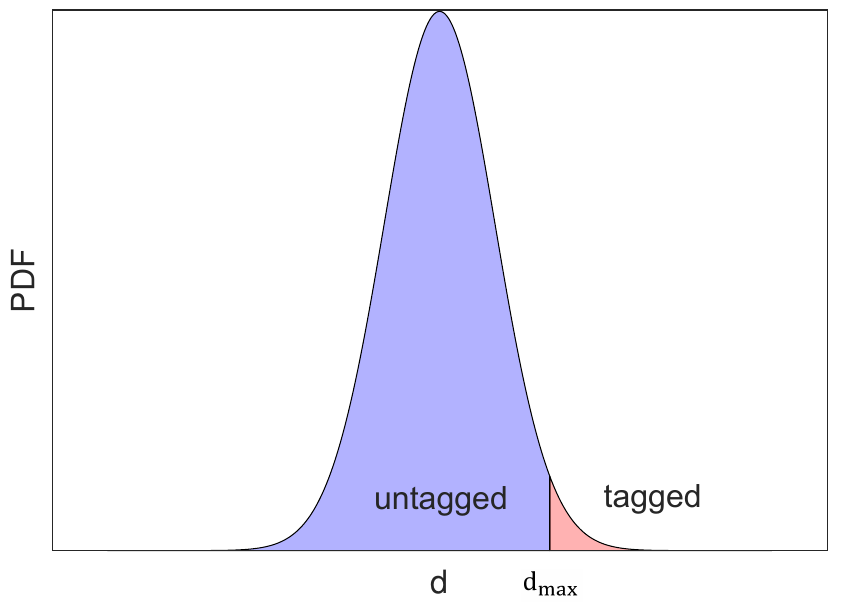}
\caption{Here, we apply a cutoff $k_{max}$ to increase the probability of untagged Gaussian states.  }
\end{figure}\label{idac}

 Note that a modified QKD protocol is needed to implement an optimal  cutoff  for CV QKD.  The modified protocol only requires  a different data recording process on the  state preparation stage while maintaining the same output states. In other words, suppose Alice desires to encode $X_{A}$ and the actual encoded data is $\sqrt{k}X_{A}$, Alice should always record the data as
  \begin{gather}\label{1}
    X_{A'}=\sqrt{k_{max}}X_{A}
\end{gather}
rather than $X_{A}$ for each pulse.

\begin{figure*}[!htb]

\begin{minipage}{0.4\linewidth}
\centerline{\includegraphics[width=1\textwidth]{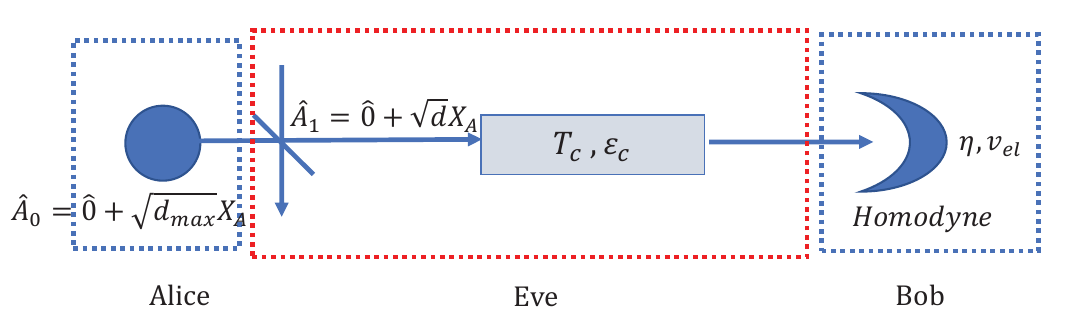}}
\centerline{(a) Untagged Gaussian state ($k\leq k_{max}$)}
\end{minipage}
\qquad
\begin{minipage}{0.4\linewidth}
\centerline{\includegraphics[width=1\textwidth]{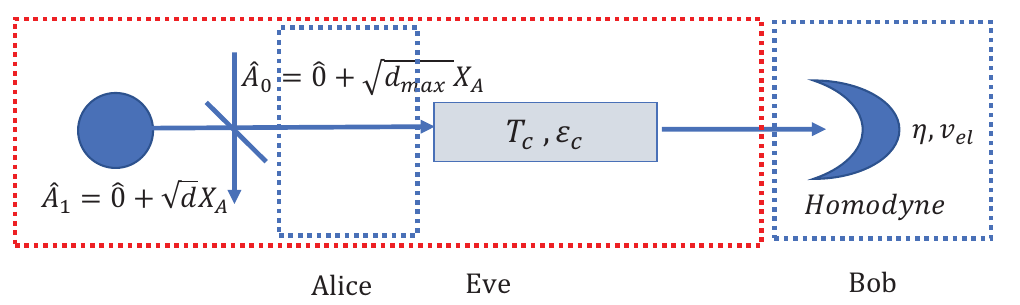}}
\centerline{(b) Tagged Gaussian state $k> k_{max}$ }
\end{minipage}

\caption{Here, we show the CVQKD system with untagged and tagged Gaussian states. Suppose that Alice always records the data as $\sqrt{k_{max}X_A}$ and  has a virtual mode $\hat{A}_0$ corresponding to the modulation $\hat{A}_0=\hat{0}+\sqrt{k_{max}}X_A $. Alice's actual output mode is $ \hat{A}_1=\hat{0}+\sqrt{k}X_A$. In Fig  6(a),  untagged states  are always secure because we conservatively assume the attenuation from a virtual mode $\hat{A}_0$ to a actual output $\hat{A}_1$  can be controlled by Eve. In Fig  6(b), tagged states   are insecure  if we consider the same attenuation mentioned before is controlled by Eve.}
\end{figure*}

Fig 6 shows the CV QKD system with untagged and tagged Gaussian states. In Fig  6(a), an untagged Gaussian state is always secure for Alice. Here, we conservatively assume the attenuation from a stronger pulse $A_0$  to a weaker pulse $A_1$ can be controlled by Eve.  In Fig  6(b), for each tagged signal, the  intensity is always  larger than the threshold value recorded by Alice. Following GLLP security proof \cite{Gottesman2002} , we conservatively assume that tagged signals are insecure.   Therefore, we only consider the secret key rate extracted from untagged Gaussian states.

Suppose that a fraction $p_s$ of the pulses emitted by the source are untagged by Eve.   The secret key  for direct reconciliation can be extracted from untagged Gaussian states at an asymptotic rate as \cite{Gottesman2002}
\begin{gather}\label{1}
  R^D_{2B} = p_sH(X_{A'})-H(X_{A'}|X_B)-\chi_{A'E,p_s}  \\
      =  I_{A'B}-(1-p_s)H(X_{A'})-\chi_{A'E,p_s}
\end{gather}
The secret key  for reverse reconciliation can be shown as
\begin{gather}\label{1}
  R^R_{2B} = p_sH(X_B)-H(X_B|X_{A'})-\chi_{BE,p_s}   \\  \notag
      = p_sH(X_B)-[H(X_B)-H(X_{A'})+H(X_{A'}|X_B)]-\chi_{BE,p_s} \\ \notag
      = I_{A'B}-(1-p_s)H(X_B)-\chi_{BE,p_s}
\end{gather}
where $X_{A'}$ and $X_B$ are Alice's and Bob's recording data, $p_sH(X_{A'})$ and $p_sH(X_{B'})$  is the  differential entropy used to generate the secret key rate depending on direct reconciliation or reverse reconciliation,  $H(X_{A'}|X_B)$ and $H(X_{B}|X_{A'})$ is the conditional differential  entropy for error correction, $\chi_{A'E,p_s}$ is the Holevo information between Alice and Eve for the untagged states, and $\chi_{BE,p_s}$ is the Holevo information between Bob and Eve for the untagged states. The Holevo information between Alice/Bob and Eve should be eliminated by the privacy amplification process. $H(X_{A'})$ and $ H(X_{A'}|X_B)$ and $H(X_B)$  can be directly estimated by Alice and Bob's data.  Given the reconciliation efficiency $\beta$, the secret key rate can be shown as
\begin{gather}\label{1}
  R^D_{2B} = \beta I_{A'B}-(1-p_s)H(X_{A'})-\chi_{A'E,p_s}\\ \notag
  R^R_{2B}  = \beta I_{A'B}-(1-p_s)H(X_B)-\chi_{BE,p_s}
\end{gather}

Next, we need to find a bound for the  Holevo information. Mathematically, it can be shown that Holevo information is monotonically increasing on the  domain of $k$. Physically, when the input pulse has a stronger intensity, Eve can obtain more information about Alice's and Bob's recorded results.  Therefore, for the untagged states, the Holevo information can be bounded
\begin{gather}\label{1}
  \chi_{BE,p_s} \leq p_s\chi_{BE}, \\    \notag
   \chi_{A'E,p_s} \leq p_s\chi_{A'E},
\end{gather}
where $\chi_{A'E}$ and $\chi_{BE}$ are the Holevo mutual information between Alice/Bob and Eve estimated from Alice's and Bob's recording results $X_{A'}$ and $X_B$.

Next, we will estimate  the equivalent transmittance $T_s$ and excess noise $ \varepsilon_s$.  According to the Supplementary Section III, the equivalent  $T_s$ and $ \varepsilon_s$  can be expressed as
\begin{gather}
T_s=<\sqrt{k}>^2/k_{max} \simeq (1-\frac{1}{8}V_k)^2/k_{max}, \\  \notag
\varepsilon_s=\frac{V_A}{T_s}-k_{max}V_A \simeq \frac{1}{4}V_AV_kk_{max},
\end{gather}

In addition to the channel transmittance $T_c$ and excess noise $\varepsilon_c$, Alice and Bob would estimate an overall transmittance $T$ and  excess noise $\varepsilon$ such that
\begin{gather}\label{1}
  T=T_sT_c,           \\  \notag
  \varepsilon=\varepsilon_c/T_s+\varepsilon_s
\end{gather}

For the secret key rate evaluation, we compare the secret key rates for two intensity fluctuation models: Gaussian distribution and uniform distribution. We still use the parameters in the Table  \ref{table1}. For the optimization, if we increase the $k_{max}$,  $p_s$ will be increased, while $T_s$ will be decreased. Therefore, we need to optimize $k_{max}$ to get the maximum secret key rates.

\begin{figure}[!htb]
\centering
\subfigure[Secret key rate vs transmission distance  ]{
\includegraphics [width=75mm,height=55mm]{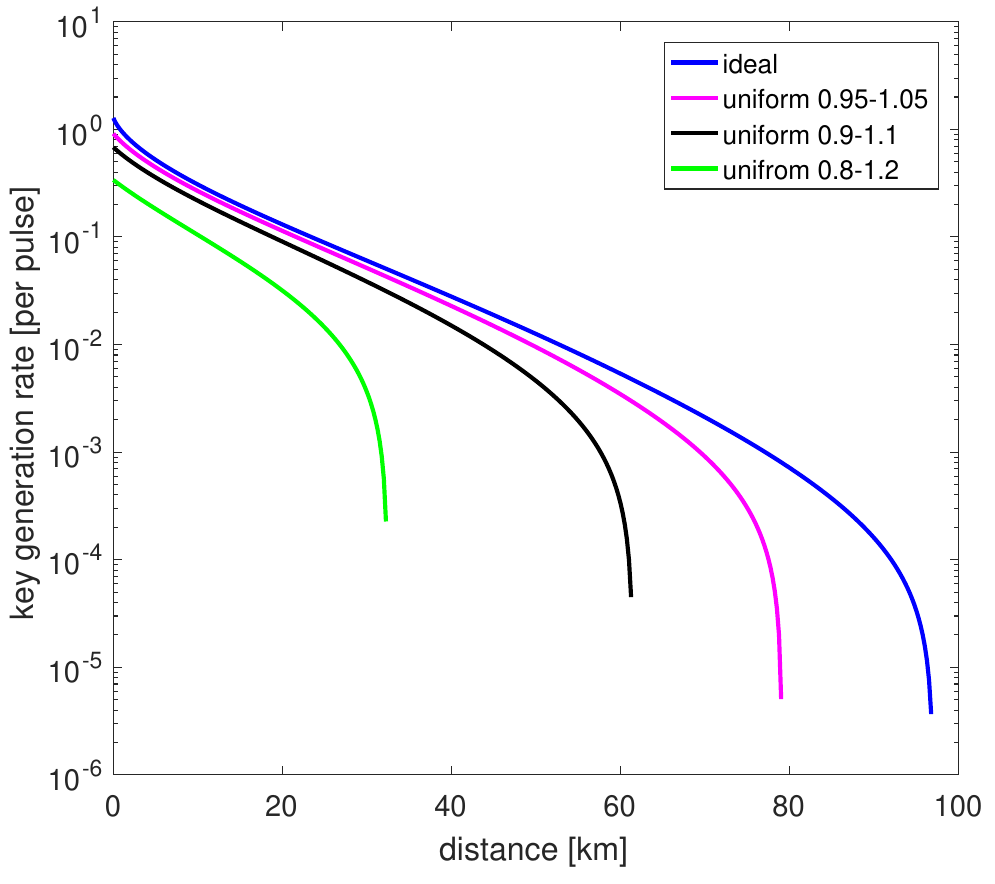}}
\subfigure[Optimal $k_{max}$ vs transmission distance]{
\includegraphics [width=75mm,height=55mm]{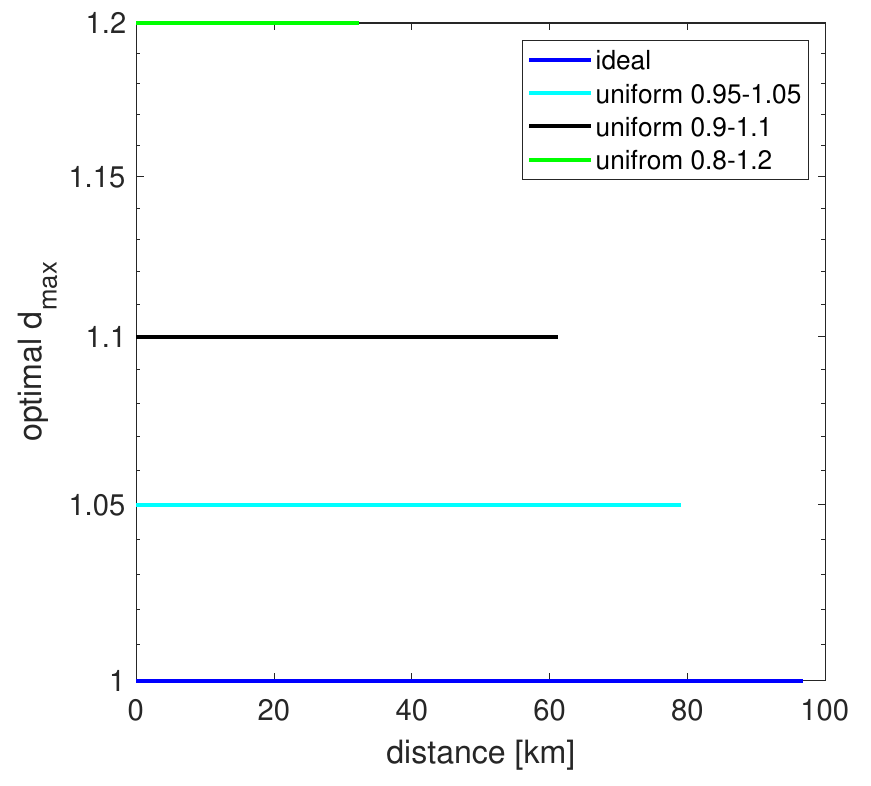}}

\caption{Here, we optimize the secret key rate $R^R_{2B}$ for uniform distribution. (a) Optimal secret key rate versus transmission distance for different uniform distributions. (b) Optimal $k_{max}$ versus transmission distance for different uniform distributions.}
\end{figure}\label{idac}

Fig 7 shows the key rate optimization results for the uniform distribution. Here, we consider the reverse reconciliation scheme. Compared to case (2A), the maximum transmission distance decreases faster due to intensity fluctuations. The maximum transmission distance will  drop by about 20 km even if  the pulse intensity  fluctuates  5\%. Meanwhile, the optimal $k_{max}$ will always be the maximum value of its domain for a uniform distribution.

\begin{figure}[!htb]
\centering
\subfigure[Secret key rate vs transmission distance  ]{
\includegraphics [width=75mm,height=55mm]{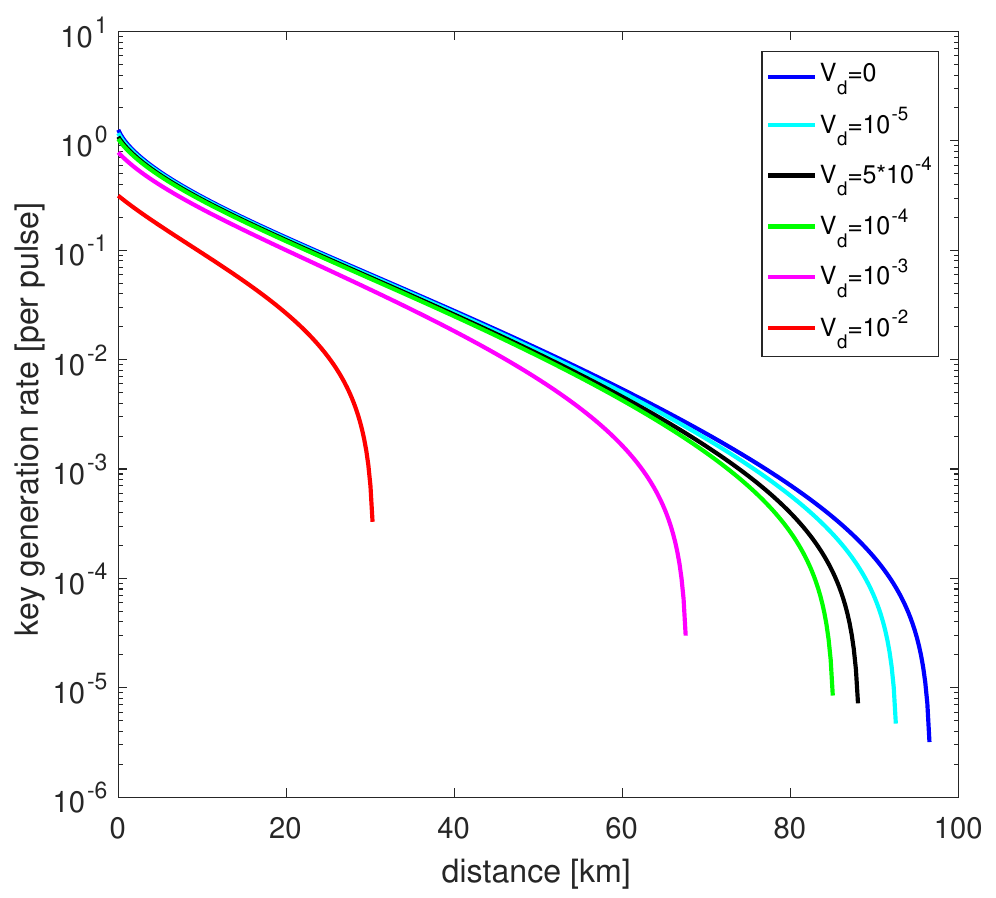}}
\subfigure[Optimal $k_{max}$ vs transmission distance]{
\includegraphics [width=75mm,height=55mm]{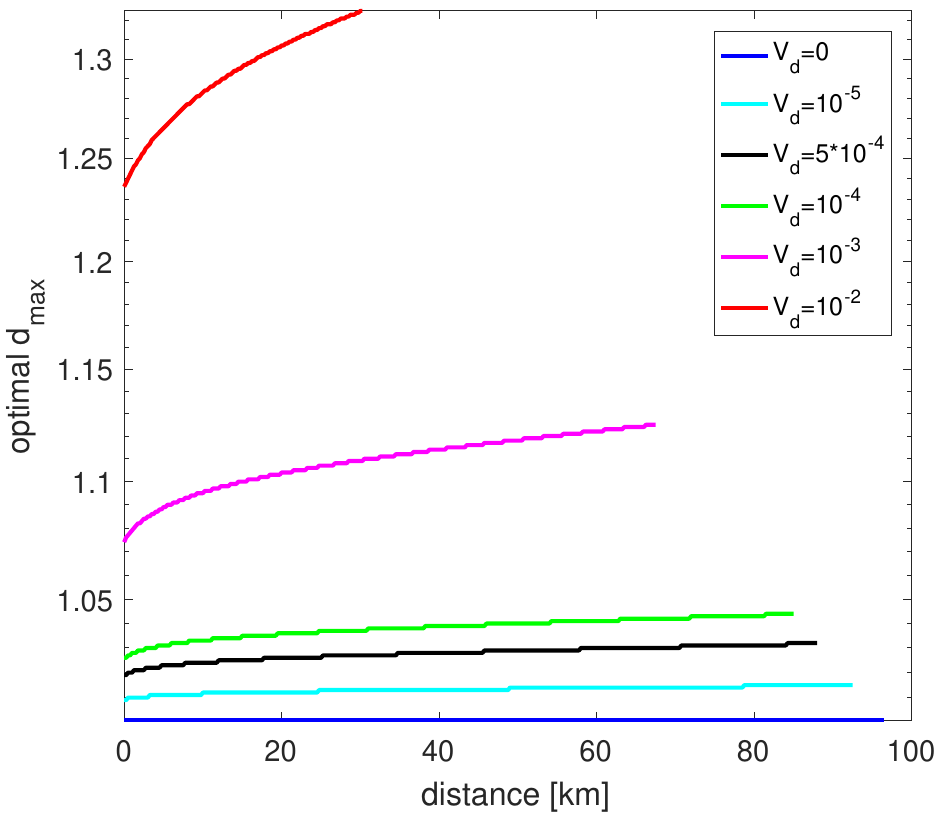}}

\caption{Here, we optimize the secret key rate $R^R_{2B}$ for uniform distribution. (a) Optimal secret key rate versus transmission distance for different uniform distribution. (b) Optimal $k_{max}$ versus transmission distance for different uniform distribution.}
\end{figure}\label{idac}

Fig 8 shows the key rate optimization results for the Gaussian distribution. Here, we also consider the reverse reconciliation scheme.  The maximum transmission distance decreases rapidly  when the intensity fluctuations increase. Other than the uniform distribution, the optimal $k_{max}$ will be monotonically increasing as a function of distance. When comparing these two intensity fluctuation models with same variance, we find that QKD with Gaussian distributed variation will have a lower key rate and transmission distance, since it always has a tail part for tagged Gaussian states.

~\\

\textbf{ Secret key rate with finite-size effects  }

\begin{figure}[!htb]
\centering
\includegraphics [width=80mm,height=60mm]{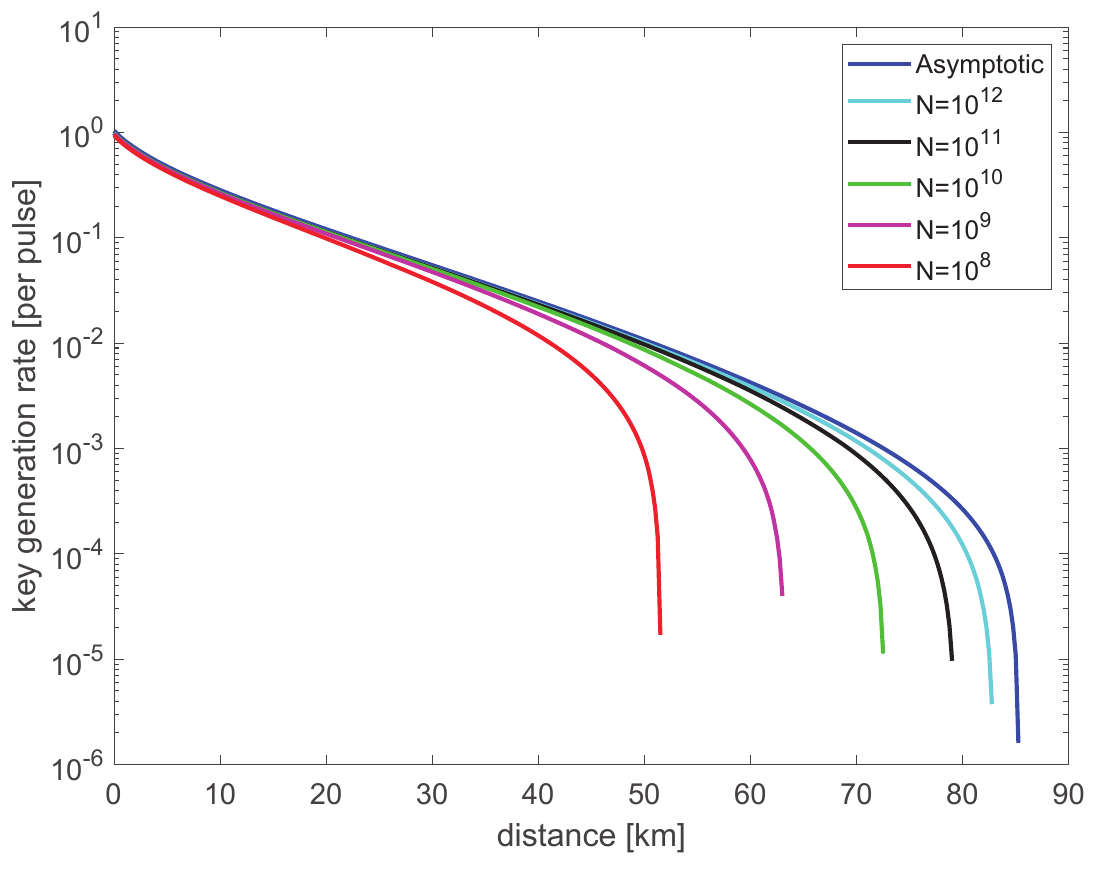}
\caption{Here, we compute the secret key rate vs distance with finite-size effects. Numerically optimized secret key rates are obtained for a fixed block size $N=10^s$ with s=8,9,10,11 and 12.  The rightmost curve corresponds to the asymptotic secret key rate. Here, we consider the Gaussian distribution model with variance $10^{-4}$.
The failure probability of parameter estimation  is $\epsilon_{PE}=10^{-10}$. The failure probability of untagged Gaussian states  is $\epsilon_{ugs}=10^{-10}$. The failure probability of  privacy amplification is  $\epsilon_{PA}=10^{-10}$.  }
\end{figure}\label{idac}

In this section, we compute the secret key rate  under finite-size scenario. Without loss of generality, we consider  case (2B) as a example.
As discussed in \cite{Leverrier2010,Ruppert2014}, by setting confidence intervals for both $T$ and $\varepsilon$, we can can obtain the lower bound of the transmittance, $T^L$, and the upper bound of the excess noise, $\varepsilon^U$. By incorporating our tagging idea, we should also obtain the lower bound of the probability, $p^L_s$, to get untagged Gaussian states.
With the three bounds, the secret key rate  with finite-size effects, $R_f$, can be shown as\cite{Leverrier2010,Ruppert2014}:
\begin{gather}\label{1}
  R_f= \frac{n}{N} \{R^R_{2B}(p_s^L, T^L, \varepsilon^U)- \triangle(n)\}
\end{gather}
where $n$ is the number of Gaussian states used for secret key transmission, $N$ is the total number of received Gaussian states  and $\triangle(n)$  is a correction term for the achievable mutual information in the finite case. The details of estimating $ p_s^L, T^L, \varepsilon^U$ and $\triangle(n)$ can be found in the supplementary.
Note that here we consider the case (2B) with reverse reconciliation, and the form of Eq.(17) can also be applied to other key rate formulas such as $R^D_{2B}$.

Fig 9 shows the secret key rate,$R^R_{2B}$, with the finite-size effects. Our method also works well for block size from $10^8$ to $10^{12}$.  For the distance less than 30 km,  there is no  distinct advantage in terms of the secret key rate
for larger block sizes, which suggests that it may not be necessary to
go to a very large block size, especially for a small distance. On the other
hand,  it is also expected that  the key rates are approaching the asymptotic limit when the block size increases.

\section{Conclusion}\label{Sec7}
We have studied the security of  CV QKD with intensity fluctuating sources.  Generally, We divide current CV QKD systems into two cases for security proof. Depending on Alice's realistic assumptions for the devices, Alice and Bob can choose different security proofs  and obtain different secret key rates. In  case (1) , Alice can monitor the intensity fluctuation value for each pulse. She can revise her data and obtain almost the same secret key rate as  what she can obtain from the ideal CV QKD systems. Furthermore, by a refined data analysis, the maximum transmission distance can be observably improved. In  case (2), depending on the devices assumptions, we also divide CV QKD systems into two subcases (2A) and (2B). In  case (2A), both Alice and Eve cannot obtain any intensity fluctuation information  of each pulse. Here, we prove the security based on Gaussian extremality. The secret key rate will decrease if  the intensity fluctuation increases.   In  case (2B), Eve could have the intensity fluctuation  information of each pulse while Alice cannot. Here, we apply the tagging idea from \cite{Gottesman2002}. We divide the signals into tagged and untagged signals, and the secret key will only be generated from untagged signals. After considering the total error correction cost and privacy amplification, the security of case (2B) can be proved. In addition, we also validate our method under finite-size regime.
Overall, our security proofs  are simple to implement without any hardware adjustment for current CVQKD systems.  In the future, we are looking for applying our methods to solve other imperfections such as phase modulation errors or atmospheric channel effects.

\section{ACKNOWLEDGEMENTS }\label{Sec7}
We acknowledge the financial support from the Natural Sciences and Engineering Research Council of Canada
(NSERC) and Huawei Technologies Canada Co., Ltd. We
also acknowledge the funding from the University of Hong Kong start-up grant

\section{Competing interest}
The authors declare that there are no competing interests.

\section{DATA AVAILABILITY}
Data sets generated and analyzed for simulation are available from the
corresponding author on request.

\section{AUTHOR CONTRIBUTIONS}
C.L., L.Q. and H.-K.L. developed the tagging idea in the CV QKD.  C.L. performed the simulations and calculations of the secret key rate. All the authors contributed to the writing of the paper.


\begin{thebibliography}{}

\bibitem[Lo(2014)]{Lo2014}
H.-K.Lo, M. Curty and K. Tamaki, Nature Photonics 8, 595-604 (2014).

\bibitem[Weedbrook(2012)]{Weedbrook2012}C. Weedbrook, S. Pirandola, R. Garcia-Patron, N. J. Cerf, T. C. Ralph, J. H. Shapiro, and S. Lloyd, Gaussian quantum information, Rev. Mod. Phys. 84, 621 (2012).

\bibitem[Diamanti(2015)]{Diamanti2015}
E. Diamanti, and A. Leverrier, Distributing Secret Keys with Quantum Continuous Variables: Principle, Security and Implementations, Entropy 17, 6072 (2015).

 \bibitem[Liao(2017)]{Liao2017}
 S-K. Liao, et al., Nature 549, 43-47 (2017).


\bibitem[Ma(2016)]{Ma2016}
 C. Ma et al., Silicon photonic transmitter for polarization-encoded quantum key distribution, Optica 3(11), 1274 (2016).

\bibitem[sibson(2017)]{Sibson2017}
 P. Sibson, et al., Integrated silicon photonics for high-speed quantum key distribution, Optica 4(2), 172 (2017).


\bibitem[Li(2017)]{Li2017}
C. Li, M. curty, F. Xu, O. Bedroya, H.-K. Lo, Secure quantum communication in the presence of phase- and polarization-dependent loss, Phys. Rev. A 98, 042324 (2018).

\bibitem[Pirandola(2017)]{Pirandola2017}
Pirandola, S., Laurenza, R., Ottaviani, C. \& Banchi, L, Fundamental limits of repeaterless quantum communications. Nat. Commun. 8, 15043 (2017).

\bibitem[Lucamarini(2018)]{Lucamarini2018}
 M. Lucamarini, Z. L. Yuan, J. F. Dynes and A. J. Shields, Overcoming the rate-distance limit of quantum key distribution without quantum repeaters, Nature 557, 400-403 (2018).

\bibitem[Xiaoqing(2019)]{Xiaoqing2019}
X. Zhong, J. Hu, M. Curty, L. Qian, and H.-K. Lo
Phys. Rev. Lett. 123, 100506 (2019).



\bibitem[zeng(2016)]{zeng2016}
Duan Huang, Peng Huang, Dakai Lin and Guihua Zeng, Long-distance continuous-variable quantum key distribution by controlling excess noise, Scientific Reports vol. 6,  19201 (2016)


\bibitem[Hong(2020)]{Hong2020}
Y. Zhang et al., Long-distance continuous-variable quantum key distribution over 202.81 km fiber, arXiv:2001.02555 [quant-ph].

 \bibitem[Lo(2012)]{Lo2012}
 H.-K. Lo, M. Curty and B. Qing, Measurement-device-independent quantum key distribution, Phys. Rev. Lett. 108, 130503 (2012).

\bibitem[Yoshino(2018)]{Yoshino2018}
K. Yoshino et. al.,  Quantum key distribution with an efficient countermeasure against correlated intensity fluctuations in optical pulses, npj Quantum Inf. 4, 8 (2018).

\bibitem[Mizutani(2019)]{Mizutani2019}
A. Mizutani et. al.,  Quantum key distribution with setting-choice-independently correlated light sources, npj Quantum Inf. 5, 8 (2019).



\bibitem[Jouguet(2013)]{Jouguet2013}
P. Jouguet, S. Kunz-Jacques, A. Leverrier, P. Grangierand E. Diamanti, Experimental demonstration of long-distance
continuous-variable quantum key distribution, Nature Photonics 7, 378-381 (2013).

\bibitem[Jouguet(2012)]{Jouguet2012}
P. Jouguet, S. Kunz-Jacques, E. Diamanti, and A. Leverrier, Analysis of imperfections in practical continuous-variable quantum key distribution, PHYSICAL REVIEW A 86, 032309 (2012).


\bibitem[Wenyuan(2017)]{Wenyuan2017}
Wenyuan Liu, Xuyang Wang, Ning Wang, Shanna Du,and Yongmin Li, Imperfect state preparation in continuous-variable quantum key distribution, PHYSICAL REVIEW A 96, 042312 (2017).



\bibitem[Filip(2008)]{Filip2008}
R. Filip, Continuous-variable quantum key distribution with noisy coherent states, PHYSICAL REVIEW A 77, 022310 (2008).

\bibitem[Usenko(2010)]{Usenko2010}
V. C. Usenko, R. Filip, Feasibility of continuous-variable quantum key distribution with noisy coherent states, PHYSICAL REVIEW A 81, 022318 (2010).



\bibitem[Shen(2011)]{Shen2011}
Y. Shen, X. Peng, J. Yang, and H. Guo, Continuous-variable quantum key distribution with Gaussian source noise, PHYSICAL REVIEW A 83, 052304 (2011)







\bibitem[Michael (2006)]{Michael2006}
M. M. Wolf, G. Giedke, and J. Ignacio Cirac, Extremality of Gaussian Quantum States, Phys. Rev. Lett. 96, 080502 (2006).

\bibitem[Raul (2006)]{Raul2006}
R. Garca-Patron and N. J. Cerf, Unconditional Optimality of Gaussian Attacks against Continuous-Variable
Quantum Key Distribution, 	Phys. Rev. Lett. 97, 190503 (2006).


\bibitem[Gottesman(2002)]{Gottesman2002}
D. Gottesman, H.-K. Lo, N. Lutkenhaus, J. Preskill, Security of quantum key distribution with imperfect devices, Quant. Inf. Comput. 5  325-360(2004).







\bibitem[Laudenbach (2018)]{Laudenbach2018}
F. Laudenbach et. al., Continuous Variable Quantum Key Distribution with Gaussian Modulation The Theory of Practical Implementations, Adv. Quantum Technol. 1800011 (2018)


\bibitem[Namiki(2018)]{Namiki2018}
Ryo Namiki, Akira Kitagawa, and Takuya Hirano,  PHYSICAL REVIEW A 98, 042319 (2018).








\bibitem[Renner(2009)]{Renner2009}
R. Renner. J.I. Cirac, de Finetti Representation Theorem for Infinite-Dimensional Quantum Systems and Applications
to Quantum Cryptography, Phys. Rev. Lett. 102, 110504 (2009).








\bibitem[Lodewyck(2007)]{Lodewyck2007}
 J. Lodewyck, et al., Quantum key distribution over 25 km with an all-fiber continuous-variable system,  Phys. Rev. A 76, 042305 (2007).


\bibitem[Devetak(2005)]{Devetak2005}
I. Devetak and A. Winter, Distillation of secret key and entanglement from quantum state, Proc. R.
Soc. Lond. A, 461, 207 (2005).

\bibitem[Jouguet(2011)]{Jouguet2011}
 P. Jouguet, S. Kunz-Jacques, and A. Leverrier, Long Distance Continuous-Variable Quantum Key Distribution with a Gaussian Modulation, Phys. Rev. A 84, 062317 (2011).







\bibitem[Jouguet(2014)]{Jouguet2014}
P. Jouguet, D. Elkouss, and S. Kunz-Jacques, Phys. Rev. A 90, 042329 (2014).




\bibitem[Leverrier(2010)]{Leverrier2010}
A. Leverrier, F. Grosshans, and Ph. Grangier, Phys. Rev. A 81,
062343 (2010).

\bibitem[Ruppert(2014)]{Ruppert2014}
L. Ruppert, V. C. Usenko, and R. Filip, Phys. Rev. A 90, 062310 (2014).





\end{thebibliography}
\end{document}